\documentclass[a4paper]{article}

\usepackage{INTERSPEECH2018}
\usepackage{subcaption}
\usepackage[hyperfootnotes=false]{hyperref}
\usepackage{enumitem}
\usepackage[hang,flushmargin]{footmisc}

\newcommand\blfootnote[1]{%
  \begingroup
  \renewcommand\thefootnote{}\footnote{#1}%
  \addtocounter{footnote}{-1}%
  \endgroup
}

\title{Conditional End-to-End Audio Transforms}
\name{Albert Haque$^{1*}$, Michelle Guo$^{1*}$, Prateek Verma$^2$}

\address{
  $^1$Department of Computer Science \quad
  $^2$Center for Computer Research in Music and Acoustics\\
  Stanford University, USA}
\email{}

\begin{document}

\maketitle

\begin{abstract}
We present an end-to-end method for transforming audio from one style to another. For the case of speech, by conditioning on speaker identities, we can train a single model to transform words spoken by multiple people into multiple target voices. For the case of music, we can specify musical instruments and achieve the same result. Architecturally, our method is a fully-differentiable sequence-to-sequence model based on convolutional and hierarchical recurrent neural networks. It is designed to capture long-term acoustic dependencies, requires minimal post-processing, and produces realistic audio transforms. Ablation studies confirm that our model can separate speaker and instrument properties from acoustic content at different context sizes. Empirically, our method achieves competitive performance on community-standard datasets.
\end{abstract}

\section{Introduction}
Humans are able to seamlessly process different audio representations despite syntactic, acoustic, and semantic variations.
Inspired by humans, modern machine translation systems often use a word-level model to aid in the translation process \cite{luong2015effective, zou2013bilingual}.
Many of these models are being used to model dependencies in music and speech for applications such as learning a latent space for representing speech, for text to speech and speech to text systems \cite{chiu2017state,wang2017tacotron}.
In the case of text-based translation, learned word vectors or one-hot embeddings are the primary means of representing natural language \cite{pennington2014glove, mikolov2013efficient}.
For speech and acoustic inputs however, word or phone embeddings are often used as a training convenience to provide multiple sources of information gradient flow to the model \cite{bourlard1990continuous, wilpon1990automatic}.
Spectrograms remain the dominant acoustic representation for both phoneme and word-level tasks since the high sampling rate and dimensionality of waveforms is difficult to model \cite{van2016wavenet}.\blfootnote{$^{*}$ Equal contribution.}

In this paper, we address the task of end-to-end audio transformations, distinct from speech recognition (speech-to-text) and speech synthesis (text-to-speech).
Specifically, we propose a fully-differentiable audio transformation model. 
Given an audio input (e.g. spoken word, single music note) conditioned on the speaker or instrument, the model learns to predict a output spectrogram for any arbitrary target speaker or instrument.
Our framework is generic and can be applied across multiple applications including timbre transfer, accent transfer, speaker morphing, audio effects, and emotion transformation.

Current models often require complex pipelines consisting of domain-specific or fine-tuned features \cite{kain1998spectral}.
By contrast, our model does not require hand-crafted pipelines and only requires \emph{conditioning} on input-output types.
We evaluate our model on two tasks: (i) transforming words spoken by a human into multiple target voices, and (ii) playing a note on an instrument and transforming the note into another instrument's while retaining the pitch.
We would like to make a key point: We do not supply the pitch variation, spectral or energy envelop to our model.
Similarly for the output audio, parameters for synthesis are not provided but are instead learned from the desired mapping. 

Our method operates on spectrograms and takes inspiration from Listen-Attend-Spell  \cite{chan2016listen} and TacoTron \cite{wang2017tacotron, shen2018is}, and is combined with ideas from natural language processing \cite{li2016persona}.
Our contributions are two fold:
\begin{enumerate}[topsep=1mm,leftmargin=4mm,itemsep=0pt]
    \item We present a fully differentiable end to end pipeline to perform audio transformations by conditioning on specific inputs and outputs for applications in speech and music.
    \item We demonstrate that our model's learned embeddings produce a meaningful feature representation directly from speech. Our model transforms a spectrogram into a target spectrogram, with the only supervision being the input speaker identity and the requested target speaker. Our method has the flexibility to capture long term dependencies present in audio.
\end{enumerate}

\subsection{Related Work}

\textbf{Voice Conversion.}
Our work is related to the problem of voice conversion (VC) \cite{abe1990voice, stylianou1998continuous}.
Several works have approached VC using statistical methods based on Gaussian mixture models, which typically involves using parallel data \cite{toda2007voice}.
Other works have also used neural network-based frameworks using restricted Boltzmann machines or feed-forward neural networks \cite{mohammadi2014voice}.
While most VC approaches require parallel source and target speech data, collecting parallel data can be expensive. 
Thus, few works have proposed parallel-data-free frameworks \cite{kinnunen2017non}. In our proposed method, time-aligned source and target speech data is not a prerequisite.

Recently, generative adversarial networks (GANs) have shown promise in image generatation and more recently in speech processing \cite{kaneko2017sequence}.
We refer the reader to the voice conversion challenge \cite{toda2016voice} for a more complete survey of VC methods. 

\textbf{Style Transfer.}
Our task of transforming audio from one style to another is closely related to the task of style transfer \cite{gatys2015texture}.
In visual style transfer \cite{gatys2015texture, johnson2016perceptual}, the computed loss is typically a linear combination of the style and content loss, ensuring that the output is semantically similar to the input, despite variations in color and texture. 
By contrast, our work directly computes the loss on the output and ground truth spectrograms.

There have been works on similar lines for audio domain.
In \cite{verma2018neural}, Verma et al. provided an additional loss term, specific for audio, to transform musical instruments.
Recent work on speech texture generation \cite{chorowski2017using} shows promise of similar ideas and techniques for learning an end to end completely differentible pipeline.
In \cite{van2017neural}, the authors introduced a version of vector quantization using variational autoencoders, which learned a code for a particular speech utterance and were able to achieve voice conversion by passing on additional speaker cues.
Developed concurrently with our work, style tokens \cite{wang2018style} and musical translation \cite{mor2018universal} show the capabilities of unsupervised learning in the audio domain.

\textbf{Text-to-Speech.} Also known as speech synthesis, text-to-speech (TTS) systems have just recently started to show promising results.
It has been shown that a pre-trained HMM combined with a sequence-to-sequence model can learn appropriate alignments \cite{wang2016first}.
Unfortunately this is not end-to-end as it predicts vocoder parameters and it is unclear how much performance is gained from the HMM aligner.
Char2wav \cite{sotelo2017char2wav} is another method, trained end-to-end on character inputs to produce audio but also predicts vocoder parameters. These models show promise of capturing semantic, speaker and word level information in an small latent space which can be used for conditioning the text in a sequence to sequence decoder. DeepVoice \cite{arik2017deep} improves on this by replacing nearly all components in a standard TTS pipeline with neural networks. While this is closer towards a fully differentiable solution, each component is trained in isolation with different optimization objectives.
\cite{chorowski2017using} showed a fully differentiable end to end speech modification pipeline but the results were not convincing in terms of audio quality and was more to show a proof of concept. 

WaveNet \cite{van2016wavenet} is a powerful generative model of audio and generates realistic output speech.
However it is slow due to audio sample-level autoregressive nature and requires domain-specific linguistic feature engineering. Our work take cues from the work done for personalization of chatbot response which condition the output of sequence to sequence models \cite{li2016persona} to achieve consistent responses. We deploy a similar strategy by conditioning on the type of audio transformation we need for the input audio. 
Also similar is Tacotron line of work \cite{wang2017tacotron, shen2018is}.
In Tacotron \cite{wang2017tacotron}, the authors move even closer to a fully differentiable system. 
The input to Tacotron \cite{wang2017tacotron} is a sequence of character embeddings and the output is a linear-scale spectrogram.
After applying Griffin-Lim phase reconstruction \cite{nawab1983signal}, the waveform is generated.

\section{Method}

Our method is a sequence-to-sequence model \cite{sutskever2014sequence} with attention. The encoder consists of a convolutional and pyramidal recurrent network \cite{koutnik2014clockwork}. The decoder is a recurrent network.

\subsection{Encoder}

\textbf{Convolutional Network.}
Modeling the full spectrogram would require unrolling of the encoder RNN for an infeasibily large number of timesteps \cite{sainath2015learning}.
Even with truncated backpropagation through time \cite{haykin2001kalman}, this would be a challenging task on large datasets.
Inspired by the Convolutional, Long Short-Term Memory Deep Neural Network (CLDNN) \cite{sainath2015learning} approach, we use a convolutional network to (i) reduce the temporal length of the input by using a learned convolutional filter bank. The stride, or hop size, controls the degree of length reduction. (ii) CNNs are good feature extractor that help the temporal unit better in modelling the longer dynamical features.

\textbf{Pyramidal Recurrent Network.} 
Inspired by the Clockwork RNN \cite{koutnik2014clockwork}, we use a pyramidal RNN to address the issue of learning from a large number of timesteps \cite{chan2016listen}.
A pyramidal RNN is the same as a standard multi-layer RNN but instead of each layer simply accepting the input from the previous layer, successively higher layers in the network only compute, or ``tick," during particular timesteps.
This allows different layers of the RNN to operate at different temporal scales.
WaveNet \cite{van2016wavenet} also controls the temporal receptive field at each layer of their network with dilated convolutions \cite{dutilleux1990implementation, yu2015multi}.
Formally, let $h_i^j$ denote the hidden state of a long short-term memory (LSTM) cell at the $i$-th timestep of the $j$-th layer: $  h_i^j = \textrm{LSTM}(h_{i-1}^j, h_i^{j-1}) $
For a pyramidal LSTM (pLSTM), the outputs from the immediately preceding layer, which contains high-resolution temporal information, are concatenated:
\begin{equation}\label{eq:pblstm}
    h_i^j = \textrm{pLSTM}\left(h_{i-1}^j, \left[ h_{2i}^{j-1}, h_{2i+1}^{j-1} \right]\right).
\end{equation}
In (\ref{eq:pblstm}), the output of a pLSTM unit is now a function of not only its previous hidden state, but also the outputs from two timesteps from the layer below.
Not only does the pyramidal RNN provide higher-level temporal features, but it also reduces the inference complexity.
Only the first layer processes each input timestep as opposed to all layers.
The input time slice into the encoder is conditioned on the speaker ID. This is done by concatenating a one-hot speaker encoding with the CNN output at each time step, before being fed into the recurrent network.

\begin{figure}[t]
	\begin{center}
		\includegraphics[width=1.0\linewidth]{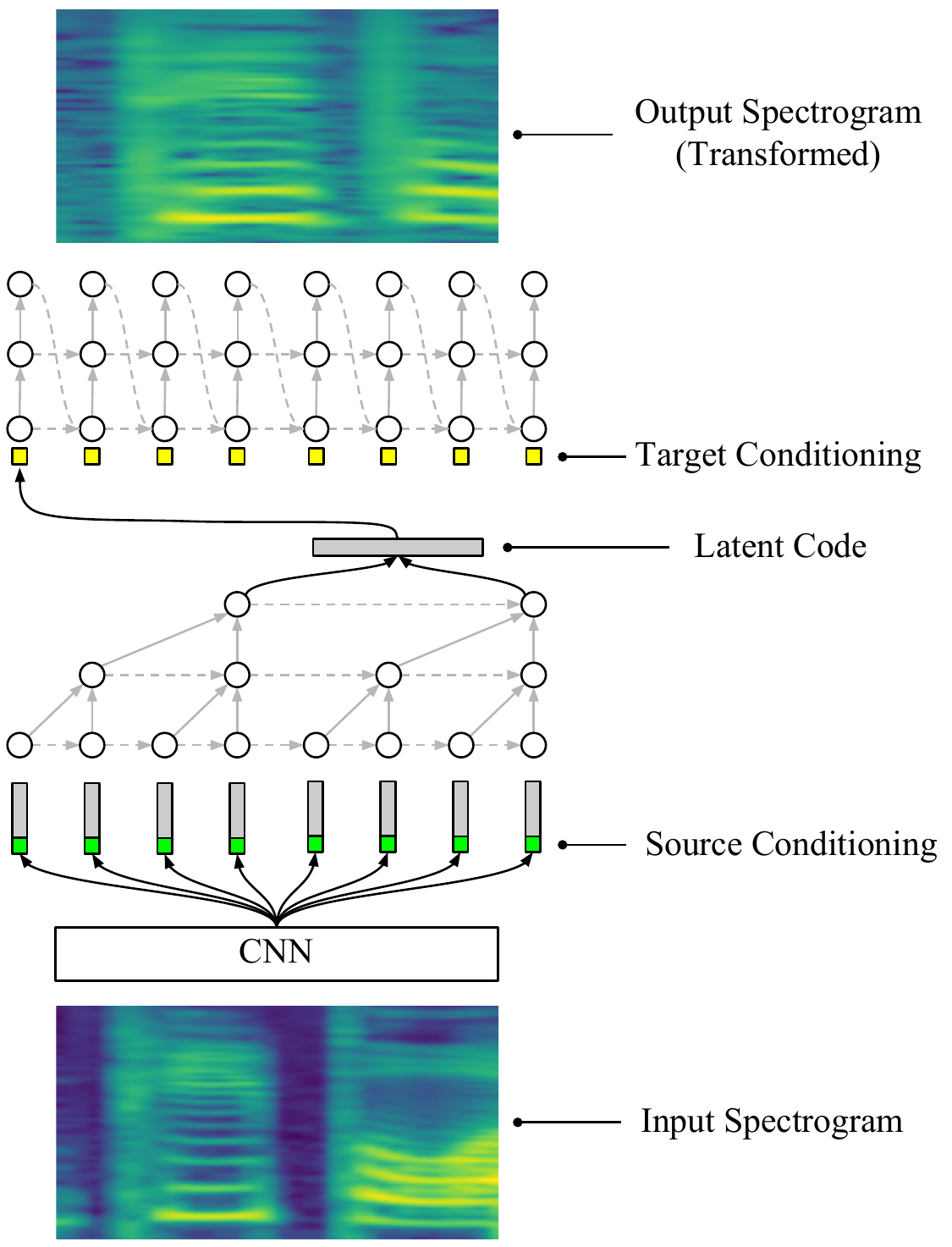}
	\end{center}
	\caption{\textbf{Overview of our model.} Yellow and green boxes denote one-hot vectors used to condition the input to each RNN. Solid lines indicate data flow. Dashed lines indicate temporal state sharing. Gray rectangles denote learned representations.
	}
	\label{fig:plot}
\end{figure}

\begin{figure*}[t]
    \centering
    \includegraphics[width=1.0\linewidth]{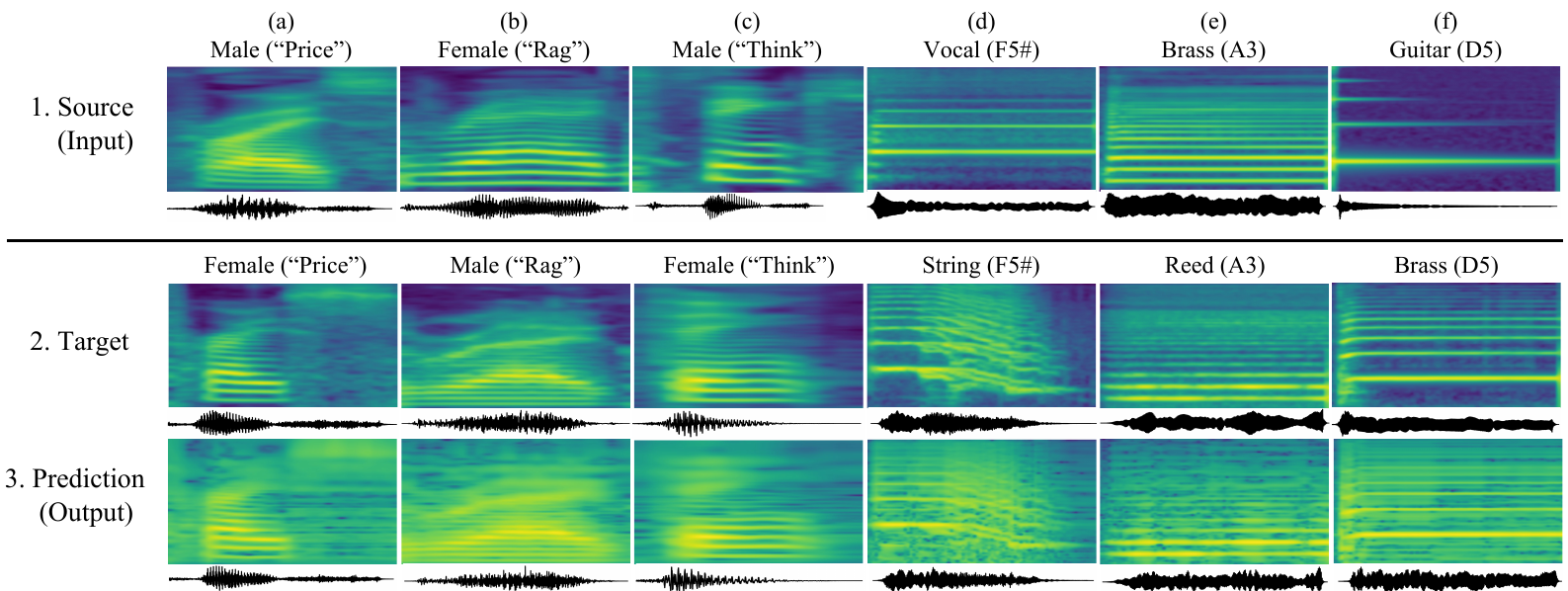}
    \caption{\textbf{Spectrogram results.} Examples on TIMIT (2a-c) and NSynth (2d-f). The speakers and instruments were present in the training set but the test set contained unseen vocabulary words or notes. The $x$ axis denotes time and the $y$ axis denotes frequency. The input, ground truth target, and our model's prediction are shown. The corresponding waveform is shown beneath each spectrogram.}
    \label{fig:spectrograms}
\end{figure*}

\subsection{Decoder}
\textbf{Attention}. Learning long-range temporal dependencies can be challenging \cite{bengio1994learning}.
To aid this process, we use an attention-based LSTM transducer \cite{chorowski2015attention, bahdanau2014neural}.
At each timestep, the transducer produces a probability distribution over the next spectrogram time-slice conditioned on all previously seen inputs.
The distribution for $y_i$ is a function of decoder state $s_i$ and context $c_i$.
The decoder state $s_i$ is a function of previous state $s_{i−1}$, previously emitted time-slice $y_{i−1}$, context $c_{i−1}$ and one hot encoding of the desired output transformation similar to \cite{li2016persona}.

The context vector $c_i$ is produced by an attention mechanism \cite{chan2016listen}.
Specifically, we define:
\begin{equation}
    \alpha_{i,j} = \frac{\exp(e_{i,j})}{\sum_{j=1}^{L} \exp(e_{i,j})}  \textrm{\quad and \quad}  c_i = \sum_u \alpha_{i,u} h_u
\end{equation}
where attention $\alpha_{i,j}$ is defined as the alignment between the current decoder time-slice $i$ and a time-slice $j$ from the encoder input.
The score between the output of the encoder (i.e., hidden states), $h_j$, and the previous state of the decoder cell, $s_{i-1}$ is computed with: $e_{i,u} = \langle \phi(s_i), \varphi(h_u)\rangle$ where $\phi$ and $\varphi$ are multi-layer perceptrons: $e_{i,j} = w^T \tanh(Ws_{i-1} + Vh_j + b)$ with learnable parameters $w$, $W$ and $V$.

\section{Experiments}

\textbf{Datasets.} We evaluate our model on two audio transformation tasks: (i) voice conversion and (ii) musical style transfer.
The TIMIT \cite{timit} and NSynth \cite{nsynth2017} datasets were used (Table \ref{tab:datasets}).
Speech examples from AudioSet \cite{audioset} were used to pre-train the model as an autoencoder.
We condition our model on the source and target style. For the case of speech, style refers to the speaker. For NSynth, it refers to the instrument type.

For all experiments, we focus on transformations on a word- or pitch-level.
This was primarily for demonstration -- adding sentence level transformations would have limited the number of training examples.
However, we note that sequence-to-sequence models are capable of encoding sentence level information in a small latent encoding vector \cite{chan2016listen}. Similarly, the decoder can model full sentences, as shown in Tacotron \cite{wang2017tacotron}.

\textbf{Audio Format.} All experiments used a sampling rate of 16 kHz with pre-emphasis of 0.97.
Audio spectrograms were computed with Hann windowing, 50 ms window size, 12.5 ms hop size, and a 2048-point Fourier transform.
Mel-spectrograms were visualized using an 80 channel mel-filterbank.

\begin{table}[t]
  \caption{\textbf{Overview of datasets.} We evaluate our method on voice conversion and musical style transfer.}
  \label{tab:datasets}
  \centering
  \begin{tabular}{l|ccc}
  \toprule
         & TIMIT \cite{timit} & NSynth \cite{nsynth2017} & AudioSet \cite{audioset} \\ \midrule
        Styles & 630 speakers & 1,006 types  & 7 classes \\
        Content  & 6,102 words & 88 pitches & ---\\
        Train & 39,834 & 289,205  & 1,010,480 \\
        Test  & 14,553 & 4,096 & --- \\ \bottomrule
    \end{tabular}
\end{table}

\textbf{Optimization.} 
The model was optimized with Adam \cite{kingma2014adam} with $\beta_1=0.9, \beta_2=0.999$ and $\epsilon=10^{-8}$.
The mean squared error was used as the loss objective.
A learning rate of $10^{-3}$ was used with an exponential decay factor of 0.99 after every epoch.
The batch size for all datasets was 128.
Models were trained for 20 epochs on NSynth and 50 epochs on TIMIT.
To train the decoder, we apply the standard maximum-likelihood training procedure, also known as teacher-forcing \cite{williams1989learning}, which has been shown to improve convergence.
The model was implemented and trained with Tensorflow on a single Nvidia V100 GPU.

\textbf{Baselines.} We evaluate three methods for audio transformations.
First, we evaluate the classical sequence-to-sequence (Seq2Seq) model \cite{sutskever2014sequence} which consists of a vanilla RNN as the encoder and a different RNN as the decoder. 
Second, we evaluate Listen, Atten, and Spell (LAS) \cite{chan2016listen}.
This is the same as the seq2seq model but the decoder is equipped with an attention mechanism that allows it to ``peek" at the encoder outputs.
Additionally they use a pyramidal RNN. Third, we evaluate a conditional sequence-to-sequence model (C-Seq2Seq). This is the same as Seq2Seq but with our conditioning mechanism.

\textbf{Evaluation Metrics.} We use subjective and objective metrics for both voice conversion and musical style transfer:
\begin{itemize}[topsep=1mm,leftmargin=4mm,itemsep=0pt]
    \item Mean opinion score (MOS), higher is better.
    \item Mel-cepstral distortion (MCD), lower is better.
    \item Side-by-side rating, higher is better.
\end{itemize}
\textit{Mel-Cepstral Distortion.} Let $y$ and $\hat{y}$ denote the ground truth and predicted mel-spectorgram, respectively. The MCD is:
\begin{equation}
    \textrm{MCD}(y, \hat{y}) = \frac{10}{\ln(10)} \sqrt{ 2\sum\limits_{t=1}^{T} ||y_t - \hat{y}_t|| }
\end{equation}
where $T$ is the number of timesteps and $t$ is the timestep slice.

\subsection{Results}

We present results on novel vocabulary words and pitches. Figure \ref{fig:spectrograms} shows results on the test set, which contain words and pitches not present in the training set. Our model is able to capture fundamental phonetic properties of each speaker or instrument and apply these properties to novel words and pitches.

\subsubsection{Mean Opinion Score}
To evaluate generative models, subjective scores \cite{shen2018is} or perceptual metrics \cite{johnson2016perceptual} are often used.
We follow this same procedure for audio generated by our model by randomly selecting a fixed set of 50 test set examples.
Audio generated from all baseline models on this set were rated by at least three normal-hearing human raters.
A total of 10 raters participated in the study and listened to the generated audio with the same over-ear headphones.
The rating scale is a 5-point numeric scale: 1. bad, 2. poor, 3. fair, 4. good, and 5. excellent.
Higher values are better.
The results of this study are shown in Table \ref{tab:sota}.
Also shown in Table \ref{tab:sota} are the mel-cepstral distances.

\begin{table}[t]
    \caption{\textbf{Comparison to existing methods.} We measure mean opinion score (MOS) and mel-cepstral distortion (MCD) on voice conversion and musical style transfer. Higher MOS is better. Lower MCD is better. The 95\% confidence interval for TIMIT MOS and MCD values are $\pm 0.024$ and $\pm 0.017$, respectively, and for NSynth, $\pm 0.016$ and $\pm 0.053$. C-Seq2Seq is a vanilla conditioned seq-to sequence model.}
    \label{tab:sota}
    \centering
    \begin{tabular}{l|cc|cc}
    \toprule
             & \multicolumn{2}{c|}{MOS} & \multicolumn{2}{c}{MCD} \\
        Method  & TIMIT & NSynth & TIMIT & NSynth \\ \midrule
        Ground Truth & 4.65 & 4.16 & --- & --- \\
        Seq2Seq \cite{sutskever2014sequence}  & 3.37 & 3.13 & 7.31 & 11.18  \\ 
        LAS \cite{chan2016listen}  & 3.52 & 3.23 & 7.40 & 11.24  \\ 
        C-Seq2Seq & 3.50 & 3.36 & 7.26 & 10.81  \\  \midrule
        Our Method  & \textbf{3.88} &  \textbf{3.43} & \textbf{6.49}  & \textbf{10.35} \\ 
        \bottomrule
    \end{tabular}
\end{table}

\subsubsection{Side-by-Side Evaluation}
We also conducted a side-by-side evaluation between audio generated by our system and the ground truth.
For each prediction-ground truth audio pair, we asked raters to give a score ranging from -1 (generated audio is worse than ground truth) to +1 (generated audio is better than ground truth).
The mean score was $-0.74 \pm 0.22$, where 0.22 denotes the 95\% confidence interval.
This indicates that raters have a preference towards the ground truth.

\subsubsection{Learned Style Representations}
\begin{figure}[t]
    \centering
    \includegraphics[width=1\linewidth]{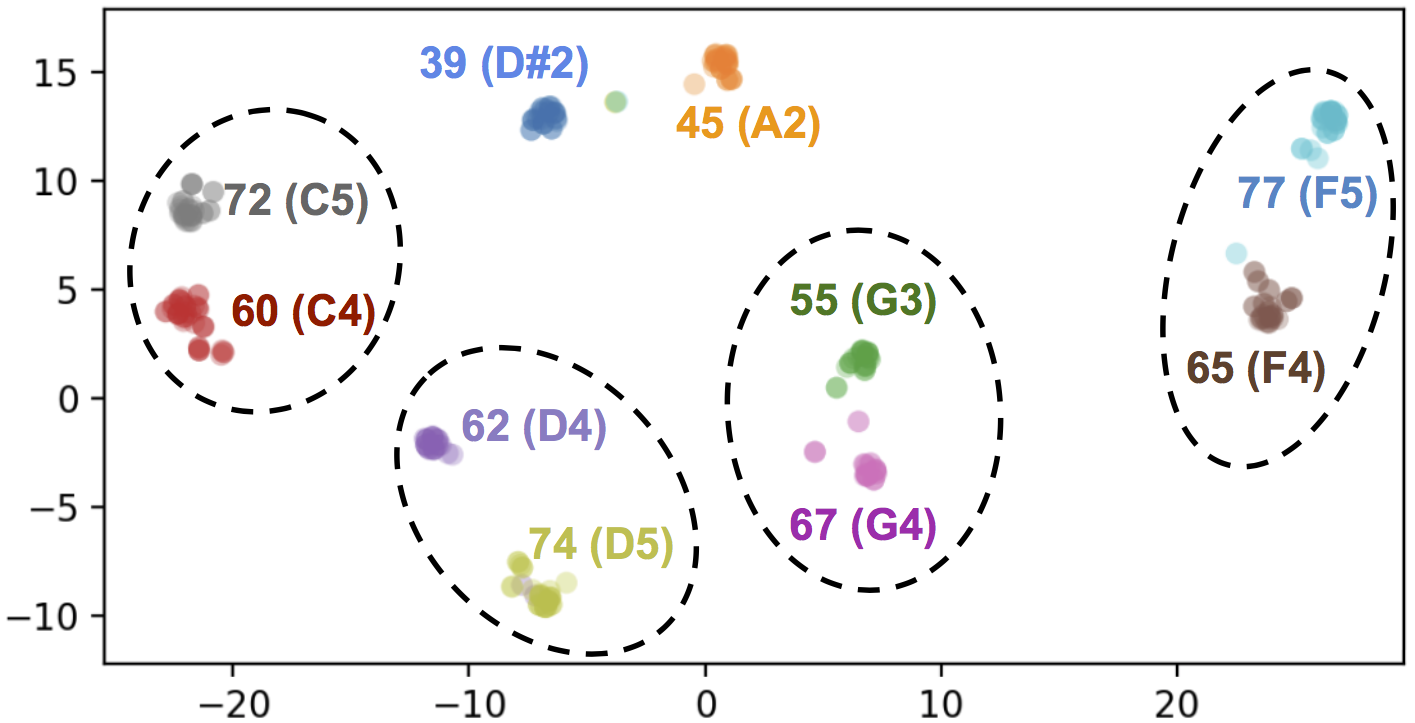}
    \caption{\textbf{Learned representations.} Each dot denotes an audio clip. Colors denote pitches. The $x$ and $y$ axes are T-SNE \cite{maaten2008visualizing} projections.  Colored text denotes the musical pitch (e.g., D\#2 refers to the note D\#, 2nd octave). Dashed circles indicate pitch classes. Despite not given any explicit pitch labels, the model was able to cluster similar musical notes.}
    \label{fig:tsne}
\end{figure}

Figure \ref{fig:tsne} shows the learned representations for the musical transformation task.
Our model can successfully cluster sounds belonging to pitch classes, rather than individual frequencies.
More interestingly, from an acoustic signals perspective, a musical octave is denoted by plus/minus 12, where 12 is the standard MIDI pitch number.
MIDI pitch 74 is one octave above MIDI 62. The resulting pitch 74 is double the frequency of pitch 62. However, pitch 74 is closer to 62 than 67, despite 67 being closer to 64 in absolute pitch.
This is because 62 and 74 have high amounts of harmonic overlap.
For audio transforms, this confirms our model can learn acoustic attributes without explicit supervision.

\subsubsection{Acoustic Context Size}

Music demonstrates different acoustic properties compared to speech \cite{gold2011speech}.
For example, speech often contains more variances in pitch whereas musical notes are held constant on the order of tens to hundreds of milliseconds.
The acoustic context size denotes the contextual window modeled by each hidden state of our network.
Larger contexts capture more acoustic content.
As shown in Table \ref{tab:receptive_field}, the context size does not significantly alter the mel-cepstral distance.
All MCD values are between 10.3 and 10.7.
Compare this to speech on TIMIT: a context size of 50 ms captures enough temporal context without overwhelming (i.e., smoothing) the hidden state.
Larger fields of size 100-200 ms produce lower quality transforms.
\blfootnote{\textbf{Acknowledgements.} We would like to thank Malcolm Slaney and Dan Jurafsky for helpful feedback. Additionally, we thank members of the Stanford AI Lab for participating in subjective experiments.}

\begin{table}[h]
    \caption{\textbf{Acoustic context size.} We varied the per timestep context size and measured MCD. Lower values are better. Plus/minus values denote the 95\% confidence interval.}
    \label{tab:receptive_field}
    \centering
    \begin{tabular}{c|cc}
    \toprule
        Context Size  & TIMIT & NSynth \\  \midrule
        12.5 ms  & $7.28 \pm 0.02$ & $\mathbf{10.35 \pm 0.05}$ \\ 
        25 ms  & $7.03 \pm 0.02$  & $10.47 \pm 0.04$ \\ 
        50 ms  & $\mathbf{6.49 \pm 0.01}$ & $10.48 \pm 0.05$ \\ 
        100 ms  & $7.11 \pm 0.02$ & $10.69 \pm 0.04$ \\ 
        200 ms & $7.32 \pm 0.02$ & $10.67 \pm 0.05$ \\
    \bottomrule
    \end{tabular}
\end{table}

\section{Conclusion}

In this work, we presented an end-to-end method for transforming audio from one style to another.
Our method is a sequence-to-sequence model consisting of convolutional and recurrent neural networks.
By conditioning on the speaker or instrument, our method is able to transform audio for unseen words and musical notes.
Subjective tests confirm the quality of our model's generated audio.
Overall, this work alleviates the need for complex audio processing pipelines and sheds new insights on the capabilities of end-to-end audio transformations.
We hope others will build on this work and extend the capabilities of end-to-end audio transforms.


\newpage
\bibliographystyle{IEEEtran}

\end{document}